\documentclass[aip,reprint,superscriptaddress,amsmath,amssymb,nofootinbib]{revtex4-2}

\usepackage{amsmath}
\usepackage{chemformula}
\usepackage[version=4]{mhchem}
\usepackage{graphicx}
\usepackage{amssymb}
\usepackage{dcolumn}
\usepackage{bm}
\usepackage{color}
\usepackage{float}
\usepackage{upgreek}
\usepackage{gensymb}
\usepackage[final]{hyperref} 
\hypersetup{
	colorlinks=true,       
	linkcolor=blue,        
	citecolor=blue,        
	filecolor=magenta,     
	urlcolor=blue
}
\usepackage{upgreek}
\usepackage[normalem]{ulem}

\newcommand{\mi}{\mathrm{i}} 

\usepackage{mathrsfs}

\usepackage{appendix}

\makeatletter

\begin{document}
\title{Clustering of chemically propelled nanomotors in chemically active environments}
	
\author{Narender Khatri}\thanks{Corresponding author: narender.khatri@utoronto.ca}
\affiliation{Chemical Physics Theory Group, Department of Chemistry, University of Toronto, Toronto, Ontario M5S 3H6,
Canada}

\author{Raymond Kapral}\thanks{Corresponding author: r.kapral@utoronto.ca}
\affiliation{Chemical Physics Theory Group, Department of Chemistry, University of Toronto, Toronto, Ontario M5S 3H6,
Canada}

\date{\today}
	
\begin{abstract}

Synthetic nanomotors powered by chemical reactions have been designed to act as vehicles for active cargo transport, drug delivery as well as a variety of other uses. Collections of such motors, acting in consort, can self-assemble to form swarms or clusters, providing opportunities for applications on various length scales. While such collective behavior has been studied when the motors move in a chemically inactive fluid environment, when the medium in which they move is a chemical network that supports complex spatial and temporal patterns, through simulation and theoretical analysis we show that collective behavior changes. Spatial patterns in the environment can guide and control motor collective states, and interactions of the motors with their environment can give rise to distinctive spatiotemporal motor patterns. The results are illustrated by studies of the motor dynamics in systems that support Turing patterns and spiral waves. This work is relevant for potential applications that involve many active nanomotors moving in complex chemical or biological environments.

\end{abstract}

\maketitle

{\bf Self-propelled active particles can serve as vehicles for cargo transport on nanoscales, and, when acting collectively, can self-assemble in various ways that have uses in materials science. These and other applications have made them targets recent research. Chemically-powered colloidal particles, like their molecular machine counterparts in biology, function out of equilibrium and take chemical energy from their environments to produce directed motion. Chemical gradients play important roles in how such active colloids self-assemble. Nonlinear chemically reacting systems share some of these features: under nonequilibrium conditions they can form complex spatial and temporal structures, like spiral waves and other chemical patterns. Here we show that these two areas of research can be combined by considering how chemically active colloidal motors self-assemble in nonlinear chemical media under nonequilibrium conditions. We find that the active colloids can use the gradients of the chemical patterns in their environments to form distinct spatial structures, or form completely new cluster assemblies and chemical patterns. Consequently, the variety of active self-assembly processes and potential applications are substantially increased in active colloidal systems with complex nonlinear chemical environments.}

\section{Introduction}

Micro- and nanoscale chemically propelled colloidal motors, operating under nonequilibrium conditions, extract chemical energy from fuel in their environment and convert it into directed motion. A considerable amount of theoretical and experimental research has been carried out on these active agents.~\cite{roadmap2020} They lie between the microscopic and macroscopic domains, where their description presents interesting theoretical challenges. Substantial experimental work has shown how these small motors can be used for diverse applications, including those related to microfluidics, active cargo transport, targeted drug delivery and other medical applications.~\cite{DWDYMS15,MHS15,Jizhuang2018,wang2019application,xu2019self,ying2019micro,soto2019onion,Tang20}

Chemically powered motors may use various mechanisms to effect propulsion,~\cite{wangbook2013} and among these phoretic mechanisms are the subject of this article.~\cite{anderson1983,golestanian2007,Kapral@JCP:2013,SenRev2013,colberg14,sanchez14,Yadav2015,OPD17,GK18a} Such motors are constructed when a particle~\cite{Howse_et_al@PRL:2007} or aggregate of particles~\cite{Ruckner_Kapral@PRL:2007,dBK13} is partially coated with a substance that catalyzes a chemical reaction and gives rise to an inhomogeneous distribution of fuel and product. The resulting self-generated concentration gradients lead to forces on the motor and fluid flows in its vicinity as a result of momentum conservation that are responsible for its propulsion. Many experiments are carried out using a metal catalyst with hydrogen peroxide fuel, but colloids with different enzyme coatings provide a way to use bio-compatible fuels for propulsion.~\cite{Dey2015,MJAHMSS15,MHPS16,ZGMS18}

Because these motors generate and respond to concentration gradients and fluid flows, they can self-organize into different inhomogeneous nonequilibrium collective states that differ from those seen in equilibrium systems. Studies of the origin and characterization of collective motor states have been carried out.~\cite{Theurkauff_et_al@PRL:2012,Thakur_Kapral@PRE:2012,Gao_et_al@JACS:2014,Cates_Tailleur@ARCMP:2015,Saha_et_al@PRE:2014,Wang_et_al@ACR:2015,Liebchen_et_al@PRL:2017,Huang_et_al@NJP:2017,Illien2017,Ginot_et_al@NC:2018,Pohl_Stark@NC:2018,Huang_et_al@JCP:2019,Stark@ACR:2018,CRRK18} These studies have shown that motor clustering processes are governed by various mechanisms ranging from simple motility-induced phase separation to clustering driven by chemotactic or hydrodynamic effects due to inhomogeneous concentration or flow fields arising from other motors in the system. In most investigations, the chemical environment in which the motors move is a simple solution of fuel and product, with possible sources and sinks for these species to maintain a nonequilibrium state. However, the chemical environment need not be simple~\cite{Bechinger_et_al@RMP:2016} and can itself be driven out-of-equilibrium to develop dynamical system states.~\cite{thakur:11_2,Robertson_Kapral@JCP:2015} If the fluid nonequilibrium state forms a chemical pattern, the chemical concentration gradients generated in this way can influence the forms that the collective behavior takes: motors can be reflected from chemical waves~\cite{thakur:11_1} or navigate chemical patterns.~\cite{CCK18}

In many biochemical situations, such as those occurring in the cell, molecular machines operate in a chemically active medium where the fuel that powers the machine's motion and its product participate in a complex nonlinear network of chemical reactions taking place under nonequilibrium conditions. The nonequilibrium states of nonlinear chemical networks are diverse and include oscillating chemical waves and stationary and dynamic patterns.~\cite{Desai_Kapral@Book:2009,Kapral_Showalter@Book:1995,Epstein1998} In this paper, we show how the collective behavior of micro- and nanoscale chemically-powered colloidal motors changes when the motors reside in and interact with a chemical network that supports such patterned system states. The results that follow show how the collective motion of motors in complex chemical environments can be modeled and, through the examples, illustrate some of the distinctive collective phenomena that occur.

\section{Motors and their reactive environment}
 \label{sec: Model}

We consider a dilute suspension of spherical active colloidal particles in a reactive multicomponent solution of chemical species in an inert solvent. Each motor with radius $R$ has a nonuniform distribution of catalytic ($C$) and noncatalytic ($N$) domains on its surface, and chemical reactions on the catalytic domains produce concentration gradients that are responsible for its propulsion through a diffusiophoretic mechanism. As is well known, sustained active motion of such particles occurs only when the system is maintained out of equilibrium by chemostats controlling fuel and product supplies to the motor, or other means. The fluid environment in which the motors move supports a nonlinear chemical network that is itself maintained in a nonequilibrium state by chemostats. Some of the chemical species in the reaction network also serve as fuel and product species that control active motor motion. The chemical dynamics in the fluid is described by reaction-diffusion equations of the form,
\begin{equation}\label{eq:RD-equation}
\partial_t c_\gamma(\bm{r},t) = D_\gamma \nabla^2 c_\gamma(\bm{r},t) + \mathcal{R}_\gamma(\bm{c}(\bm{r},t)),
\end{equation}
where $\mathcal{R}_\gamma$ is the reaction term for species $\gamma=A,B,\dots$, and $\bm{c}=(c_A,c_B, \dots)$ is used to denote the collection of all reactive solute species. The diffusion coefficient of species $\gamma$ is $D_\gamma$, and the solution is assumed to be dilute so that cross-diffusion effects can be neglected. A subset of these species, designated by the index $k$, is involved in reactions on the motor surface.

The active translational and rotational motion of a single motor can be computed using the solution of the reaction-diffusion equations subject to reactive boundary conditions on the motor surface and the presence of concentration gradients in the system.~\cite{Gaspard_Kapral@Research:2020} Here we consider the simple reaction $A \rightleftharpoons B$ on the motor, so $k=A,B$ denotes the fuel and product. In this case, the diffusiophoretic linear and angular velocities are given by $\bm{V}_d=\sum_k(\zeta_k  \bar{c}_k \bm{u} -\xi_k  \bm{g}_k)$ and $\Omega_d= \sum_k \lambda_k \bm{u} \wedge \bm{g}_k$, respectively, where $\bm{u}$ is the orientation vector of the motor, $\bar{c}_k$ is the concentration of species $k$ extrapolated to the motor center, and $\bm{g}_k$ is the gradient of species $k$. The first term in $\bm{V}_d$ is due to the propulsion of the motor by its self-generated concentration gradient, and the form in the equation involving the constant prefactor $\zeta_k$ and the concentration $\bar{c}_k$ far from the motor results from the evaluation of the surface average of this concentration gradient.

The situation is more complicated for a system containing many motors since each motor's motion depends on the instantaneous inhomogeneous concentration fields produced by all other motors in the system. Approximate mean field equations of motion have been constructed~\cite{Saha_et_al@PRE:2014,SMBL15,Liebchen_et_al@PRL:2017,Pohl_Stark@NC:2018,Tang20} that are often based on moments of the motor distribution function $f(\bm{r},\bm{u},t)$: density, $n_C(\bm{r},t)=\int d^2 u\; f(\bm{r},\bm{u},t)$, polarization density, $\bm{p}(\bm{r},t)=\int d^2 u\;\bm{u}\; f(\bm{r},\bm{u},t)$, etc.  We use the evolution equation for the local colloid density that takes the form (see Eq.~(78) in Ref.~\cite{Gaspard_Kapral@Research:2020}),
\begin{equation}
\partial_t n_{C} = \boldsymbol{\nabla} \biggl\{ D_{t}^{\rm eff} \boldsymbol{\nabla} n_{C} - n_{C} \sum_{k} \mathcal{M}_{k} \boldsymbol{\nabla} c_{k} \biggr \},\label{eq:Density_nC}
\end{equation}
where
\begin{equation}\label{eq: M_expression}
 \mathcal{M}_{k} = \xi_{k} + \frac{V_{\rm sd}}{6 D_{r}} (2 \lambda_{k} - \zeta_{k}).
\end{equation}
Here the self-diffusiophoretic velocity is $V_{\rm sd} \equiv \sum_{k} \zeta_{k} c_{k}$ and the effective translational diffusion coefficient of a motor is $D_{t}^{\rm eff} =  D_{t} + \frac{V_{\rm sd}^{2}}{6 D_{r}}$, with the translational and rotational diffusion constants given by $D_{t}$ and $D_{r}$, respectively, related to the translational and rotational friction coefficients by the Einstein relations: $D_{t} = k_{B} T/\gamma_{t}$ and $D_{r} = k_{B} T/\gamma_{r}$. We see that the motor velocity $V_{\rm sd}$ enters the motor effective diffusion coefficient and the $\mathcal{M}_{k}$ parameters. This equation is valid for dilute colloid suspensions and assumes that the rotational diffusion coefficient $D_r$ is sufficiently large that the polarization density is driven by the colloid density and species concentration gradients. For micromotors in solution with radius $R$, low Peclet number conditions apply where $Pe= V_d R/D_\gamma \ll 1$, so that advective terms in the concentration field equations may be neglected and their dynamics is described by the reaction-diffusion equations~(\ref{eq:RD-equation}).

The set of Eqs.~(\ref{eq:RD-equation})-(\ref{eq:Density_nC}) must be solved to determine the nonlinear dynamics of the system. The parameters $\zeta_k$, $\xi_k$ and $\lambda_k$ that enter into this equation are available from single motor solutions as indicated above~\cite{Gaspard_Kapral@Research:2020} and play important roles in determining the nature of the collective behavior.

\section{Motors in media with chemical waves and patterns}
 \label{sec: Model_Sec2}

 The phenomena seen in spatially-distributed far-from-equilibrium nonlinear chemical systems are extensive and often take the form of complicated spatial patterns or chemical waves.~\cite{Kapral_Showalter@Book:1995,Epstein1998,Desai_Kapral@Book:2009} The chemical gradients that accompany such nonequilibrium chemical structures can drive motor dynamics, and it is interesting to investigate the collective dynamics of motors in such media. For this purpose, we select a well-studied chemical network that can self-organize to form complex spatiotemporal structures: the Selkov model that was proposed originally as a simple scheme to describe the oscillations observed for adenosine diphosphate (ADP) and
adenosine triphosphate (ATP) in the enzymatic conversion of fructose-6-phosphate to fructose-1,6-bisphosphate in glycolysis.~\cite{Selkov@EJB:1968} The chemical network that specifies the reversible reaction dynamics of this system may be written as~\cite{Richter_et_al@PTP:1981}
\begin{align}\label{eq: Selkov_R_Scheme}
  \ce{$F$ &<=>[$k_1$][$k_{-1}$] $A$}, \nonumber \\ 
  \ce{$A$ + $2B$ &<=>[$k_2$][$k_{-2}$] $3B$},\\ 
  \ce{$B$ &<=>[$k_3$][$k_{-3}$] $G$}. \nonumber 
\end{align}
In this model $A$ and $B$ correspond to ATP and ADP, respectively. The fluid phase species $F$ and $G$ participate in reactions that produce $A$ (ATP) from $F$ and degrade the product $B$ (ADP) to $G$, which are simplified versions of the corresponding more complex biological processes that carry out these functions in the cell. These molecular species have constant concentrations, and maintain the system in a nonequilibrium state. In addition to these reactions that occur in the fluid phase, reactions occur on motor catalytic domains, so we have the following additional contribution to the mechanism:
\begin{equation} \label{eq: Reaction}
\ce{$A$ + $C$ ->[$\kappa_+$] $B$ + $C$},
\end{equation}
where $\kappa_+=k_+ /(4\pi R^2)$ is the forward rate constant per unit surface area of the motor, and, for simplicity, we assumed that the motor reaction is irreversible. The corresponding reaction terms are:
\begin{eqnarray}
\mathcal{R}_{A} &=& k_{1} - k_{-1} c_{A} - k_{2} c_{A} c_{B}^{2} + k_{-2}c_{B}^{3} - k_{+} c_{A}n_C,
\label{eq: Exp_RA}\\
\mathcal{R}_{B} &=& k_{-3} - k_{3} c_{B} + k_{2} c_{A} c_{B}^{2} - k_{-2}c_{B}^{3} + k_{+} c_{A}n_C, \label{eq: Exp_RB}
\end{eqnarray}
where the constant concentrations of $F$ and $G$ feed species are incorporated in $k_{1}$ and $k_{-3}$, respectively.

It is convenient to use a dimensionless description~\cite{Richter_et_al@PTP:1981} where all length variables are scaled by $l_{0} = \sqrt{D_{A}/(h k_{3})}$, $\boldsymbol{r'} = \boldsymbol{r}/l_{0}$, time by $\tau = 1/(h k_{3})$, $t' = t/\tau$, and concentrations of species $c_{k}$ by $c_{0} = \sqrt{h k_{3}/k_{2}}$, $c_{k}' = c_{k}/c_{0}$, where $h$ is a dimensionless constant used to adjust the time scale. Furthermore, the global density of motors is a constant $n_C^0$, and we scale the densities $n_C$ by this constant, $n_C' = n_C/n_C^0$. Taking $k_{2} = k_{-2}$, in dimensionless variables, the coupled Eqs.~(\ref{eq:RD-equation})-(\ref{eq:Density_nC}) read
\begin{align}
\partial_t c_{A} &=  \nabla^{2} c_{A} + \mathcal{R}_{A}, \label{eq: Dimensionless_RD_CA} \\
\partial_t c_{B} &=   r_{BA} \nabla^{2} c_{B}  + \mathcal{R}_{B},  \label{eq: Dimensionless_RD_CB} \\
\partial_t n_C &= \boldsymbol{\nabla} \biggl\{ \widetilde{D}_{t}^{\rm eff} \boldsymbol{\nabla} n_C - n_C \sum_{k} \mathcal{\widetilde{M}}_{k} \boldsymbol{\nabla} c_{k} \biggr \}, \label{eq: Dimensionless_Density_nC}
\end{align}
with
\begin{align}
\mathcal{R}_{A} &= g_{1} - f c_{A} - c_{A} c_{B}^{2} + c_{B}^{3} - \widetilde{k}_{+}  n_C c_{A}, \label{eq: Dimensionless_Exp_RA}\\
\mathcal{R}_{B} &= g_{2} - \frac{c_{B}}{h} + c_{A} c_{B}^{2} - c_{B}^{3} + \widetilde{k}_{+} n_C c_{A}, \label{eq: Dimensionless_Exp_RB}
\end{align}
and we dispensed with the primes for better readability. Here the dimensionless parameters are given by $g_{1} = k_{1}/(h c_0 k_{3}),\; f = k_{-1}/(h k_{3}),\;  g_{2} = k_{-3}/(h c_0 k_{3}),\; \widetilde{k}_{+} = k_{+} n_C^0/(h k_{3}),\; r_{BA} = D_{B}/D_{A},\; \widetilde{D}_{t} = D_{t}/D_{A},\; \widetilde{D}_{t}^{\rm eff} = D_{t}^{\rm eff}/D_{A}$, and $\mathcal{\widetilde{M}}_{k}= \mathcal{M}_{k} c_{0}/D_{A}$.

The reaction-diffusion system for the Selkov model can support a variety of homogeneous and inhomogeneous nonequilibrium states. We focus now on two such states: Turing patterns and spiral waves.

\section{Turing patterns} \label{sec:turing}

A Turing pattern is formed when a steady state that is stable to homogeneous perturbations is unstable to spatially inhomogeneous perturbations under specified conditions. While Turing patterns were first discussed in the context of morphogenesis,~\cite{Turing:1952} they have been observed in laboratory experiments on chemical systems.~\cite{Castets_et_al@PRL:1990,ouyang1991} To review the conditions for a Turing bifurcation,~\cite{Desai_Kapral@Book:2009} we consider first the bifurcation structure of the homogeneous system where spatial gradient terms are removed from Eqs.~(\ref{eq: Dimensionless_RD_CA})-(\ref{eq: Dimensionless_Density_nC}). Since the motor density is conserved, Eq.~(\ref{eq: Dimensionless_Density_nC}) plays no role, and we set the (scaled) density to $n_C=1$ in Eqs.~(\ref{eq: Dimensionless_Exp_RA}) and (\ref{eq: Dimensionless_Exp_RB}). The resulting set of ordinary differential equations is
\begin{equation}\label{eq:RD-hom}
\frac{d }{dt}\bm{c}(t)= \bm{\mathcal{R}}(\bm{c}(t)),
\end{equation}
where $\bm{c}=(c_A,c_B)$ and $\bm{\mathcal{R}}=(\mathcal{R}_A,\mathcal{R}_B)$. The steady state solutions $\bm{c}^0$ of this equation satisfy $\bm{\mathcal{R}}(\bm{c}^0)=0$, and their stability to homogeneous perturbations can be determined from a linear stability analysis of Eq.~(\ref{eq:RD-hom}). This calculation shows that the solutions will be stable to such perturbations if ${\rm Tr} \;\bm{{\sf R}} < 0$ and ${\rm det} \;\bm{{\sf R}}>0$, where $\bm{{\sf R}}=(\partial \bm{\mathcal{R}}/ \partial\bm{c})|_{\bm{c}^0}$. We choose kinetic parameters to satisfy these conditions: $g_{1} = 0.5,\; g_{2}= 0.06, \; f = 0.1, \; h = 1$  and $\widetilde{k}_+ \le 0.0001$.~\cite{Note:Hopf}

The stability of the steady state solutions to spatially inhomogeneous perturbations is determined from a linear stability analysis of the coupled system~(\ref{eq: Dimensionless_RD_CA})-(\ref{eq: Dimensionless_Density_nC}). Writing $\bm{\mathcal{C}}=(\bm{c},n_C)$, $\bm{\mathcal{C}}^0=(\bm{c}^0,1)$ and $\delta \bm{\mathcal{C}}= \bm{\mathcal{C}}-\bm{\mathcal{C}}^0$ we obtain linearized evolution equation,
\begin{equation}
\partial_t \delta \bm{\mathcal{C}}(\bm{r},t) =\widehat{\bm{\mathsf{L}}}\; \delta \bm{\mathcal{C}}(\bm{r},t),
\end{equation}
with
\begin{equation}\label{eq: Linearized_Equations}
\begin{aligned}
\widehat{\bm{\mathsf{L}}}=
&\begin{pmatrix}
 \nabla^{2} + \mathsf{R}_{AA} &  \mathsf{R}_{AB} & \mathsf{R}_{AC}\\
\mathsf{R}_{BA} &  r_{BA} \nabla^{2} + \mathsf{R}_{BB} & - \mathsf{R}_{AC}\\
-\mathcal{\widetilde{M}}_{A}\nabla^{2} & -\mathcal{\widetilde{M}}_{B}\nabla^{2} &  \widetilde{D}_{t}^{\rm eff}\nabla^{2}
\end{pmatrix},\\
\end{aligned}
\end{equation}
where now $\bm{{\sf R}}=(\partial \bm{\mathcal{R}}/ \partial\bm{\mathcal{C}})|_{\bm{\mathcal{C}}^0}$. For perturbations of the form $\delta \mathcal{C} \sim e^{i \bm{q} \cdot \bm{r} + s t}$, the decay constants $s(q)$ can be found from the secular equation $\mid s(q) \bm{1} -\bm{\mathsf{L}}(q) \mid =0$, where $\bm{1}$ is the unit matrix and $\bm{\mathsf{L}}(q)$ is obtained from $\widehat{\bm{\mathsf{L}}}$ by the replacement $\nabla^2 \to -q^2$. We denote the three solutions of the secular equation by $s_i(q)$, $(i=1,2,3)$, and the corresponding eigenvectors as $V_i(q)= \sum_{j=A}^C \alpha_{ij}(q) \delta \mathcal{C}_j(q)$, where the coefficients $\alpha_{ij}(q)$ follow from the diagonalization of $\bm{\mathsf{L}}(q)$.

It is instructive to consider the simple situation where $\widetilde{k}_+=0$. For this case, $\mathsf{R}_{AC}=0$ so that the Selkov concentrations $\bm{c}$ no longer depend on the motor density $n_C$. Correspondingly, in the motor density equation, we have the replacements $\widetilde{D}_{t}^{\rm eff} \to \widetilde{D}_t$ and $\widetilde{\mathcal{{M}}}_{k} \to \widetilde{\xi}_k$. The latter replacement shows that the colloids can respond to concentration gradients in the system but they are not self-propelled. The secular equation factors into motor and Selkov reaction parts. The motor mode has eigenvalue $s(q)=-\widetilde{D}_{t}q^2$, so that it is stable, while the Selkov eigenvalues are given by the solution to
\begin{equation}\label{eq: Secular_equation_No_Reaction}
(s +  q^{2} - \mathsf{R}_{AA})(s +  r_{BA} q^{2} - \mathsf{R}_{BB})- \mathsf{R}_{AB} \mathsf{R}_{BA}=0.
\end{equation}
A necessary condition for a Turing instability is $r_{BA} \mathsf{R}_{AA} +\mathsf{R}_{BB} >0$, which implies that the diffusion coefficient ratio, $r_{BA}=D_B/D_A \ne 1$, since ${\rm Tr} \;\bm{{\sf R}}=\mathsf{R}_{AA} +\mathsf{R}_{BB} < 0$. For the chosen kinetic parameters given above, we have $\mathsf{R}_{AA} = -0.118, \mathsf{R}_{AB} = -1.092, \mathsf{R}_{BA} = 0.018$, and $\mathsf{R}_{BB} = 0.092$, which satisfy the Turing bifurcation conditions. The onset of the Turing bifurcation occurs at $r_{BA}=0.15$ and the non-zero critical wavenumber is given by $q_c^2=(r_{BA} \mathsf{R}_{AA}  + \mathsf{R}_{BB})/(2 r_{BA}) = 0.25$.

\begin{figure}[htb!]
\centering
\resizebox{1.0\columnwidth}{!}{%
\includegraphics[scale = 1.0]{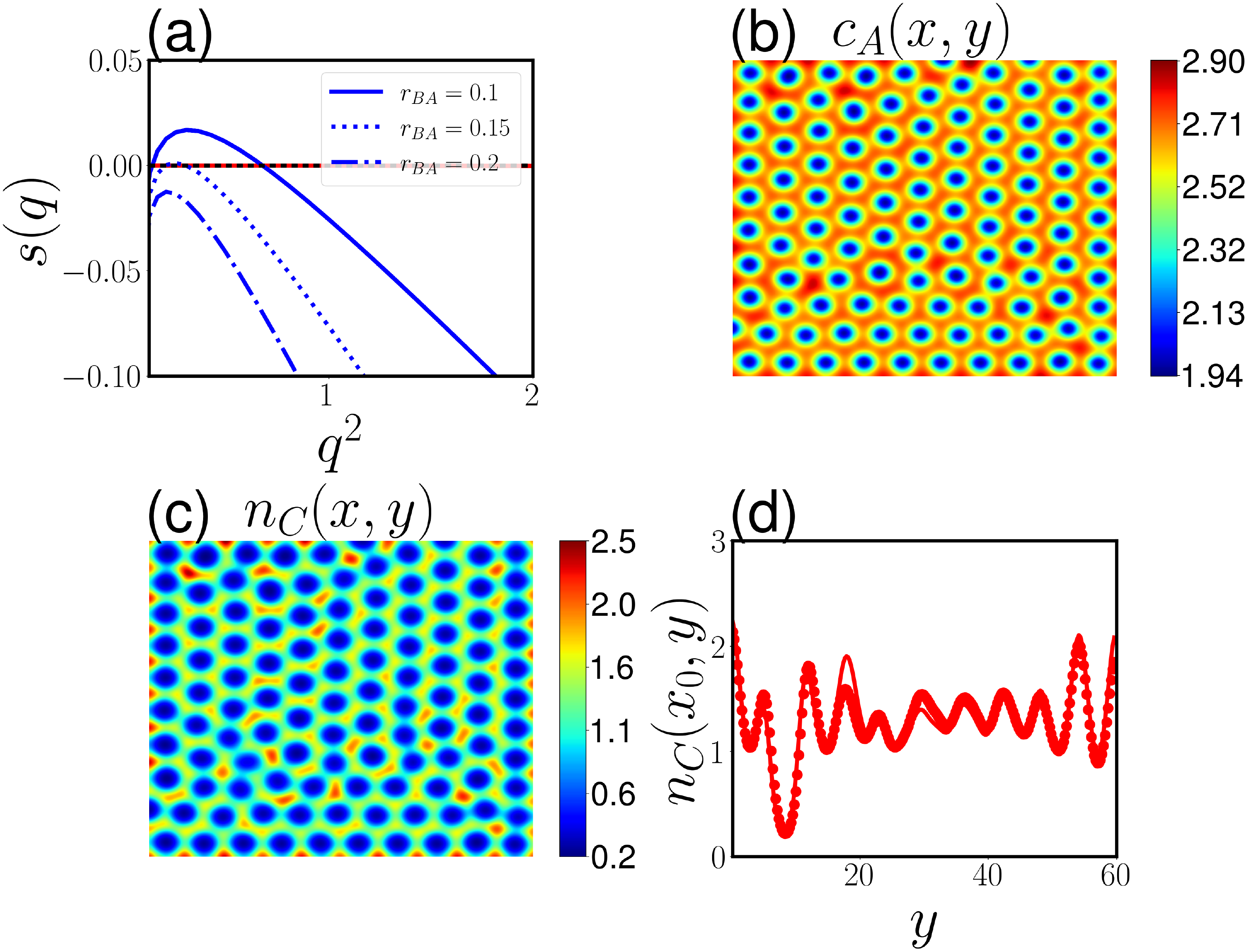}}
\caption{(a) The plot of the eigenvalue $s(q)$ that becomes unstable versus $q^2$ for $r_{BA} = 0.1, \; 0.15, \; \mathrm{and} \; 0.2$, solid, dotted, and dash-dotted blue lines, respectively. (b) The plot of the simulated concentration field of fuel species $c_A(\bm{r})$ in the form of a stationary spot Turing pattern. (c) The inhomogeneous motor density field $n_C(\bm{r})$. (d) Comparison of the simulation results (represented by symbols) with Eq.~(\ref{eq:nc-theory}) (solid line). The set parameters are: $\widetilde{k}_{+} = 0$, $\widetilde{D}_{t} = 0.0001$, and $(\mathcal{\widetilde{M}}_{A}, \mathcal{\widetilde{M}}_{B}) = (0.004, 0.001)$. The elements of $\bm{{\sf R}}$ are $\mathsf{R}_{AA} = -0.118, \mathsf{R}_{AB} = -1.092, \mathsf{R}_{BA} = 0.018$, and $\mathsf{R}_{BB} = 0.092$. }
\label{fig:SA_Turing_Pattern}
\end{figure}
The $s(q)$ eigenvalue that becomes unstable is plotted as a function of $q^2$ in Fig.~\ref{fig:SA_Turing_Pattern} (a) for several values of $r_{BA}$ that are above and below the threshold for the Turing bifurcation. The $r_{BA}=0.1$ curve shows a band of unstable wavenumbers with a maximum in $s(q)$ close to $q_c$. To determine the chemical pattern that is selected, direct numerical simulations of the coupled Eqs.~(\ref{eq: Dimensionless_RD_CA})-(\ref{eq: Dimensionless_Density_nC}) on a quasi-two-dimensional square domain of area $L^2$ were carried out using an explicit Euler finite difference method with no-flux boundary conditions.~\cite{Hull_White@JFQA:1990} Panel (b) of the figure shows that the ${c}_A(\bm{r})$ concentration field adopts the form of a stationary pattern of spots, while panel (c) for $n_C(\bm{r})$ shows that clustering of motors has also taken place, driven by the concentration gradients generated in the Turing bifurcation (see movie S1 in the
Supplementary Material). Since the $\bm{c}(\bm{r})$ are determined solely by the Selkov kinetics, independent of the motors, the steady state motor distribution can be determined from Eq.~(\ref{eq: Dimensionless_Density_nC}) and is given by
\begin{equation}\label{eq:nc-theory}
n_C(\bm{r})= n_C(\bm{r}_0) e^{\widetilde{\bm{\xi}} \cdot(\bm{c}(\bm{r})-\bm{c}(\bm{r}_0))/\widetilde{D}_t},
\end{equation}
where $\bm{r}_0$ is a reference point in the system. This formula is compared to the simulation results in Fig.~\ref{fig:SA_Turing_Pattern} (d).

\begin{figure}[htb!]
\centering
\resizebox{1.0\columnwidth}{!}{%
\includegraphics[scale = 1.0]{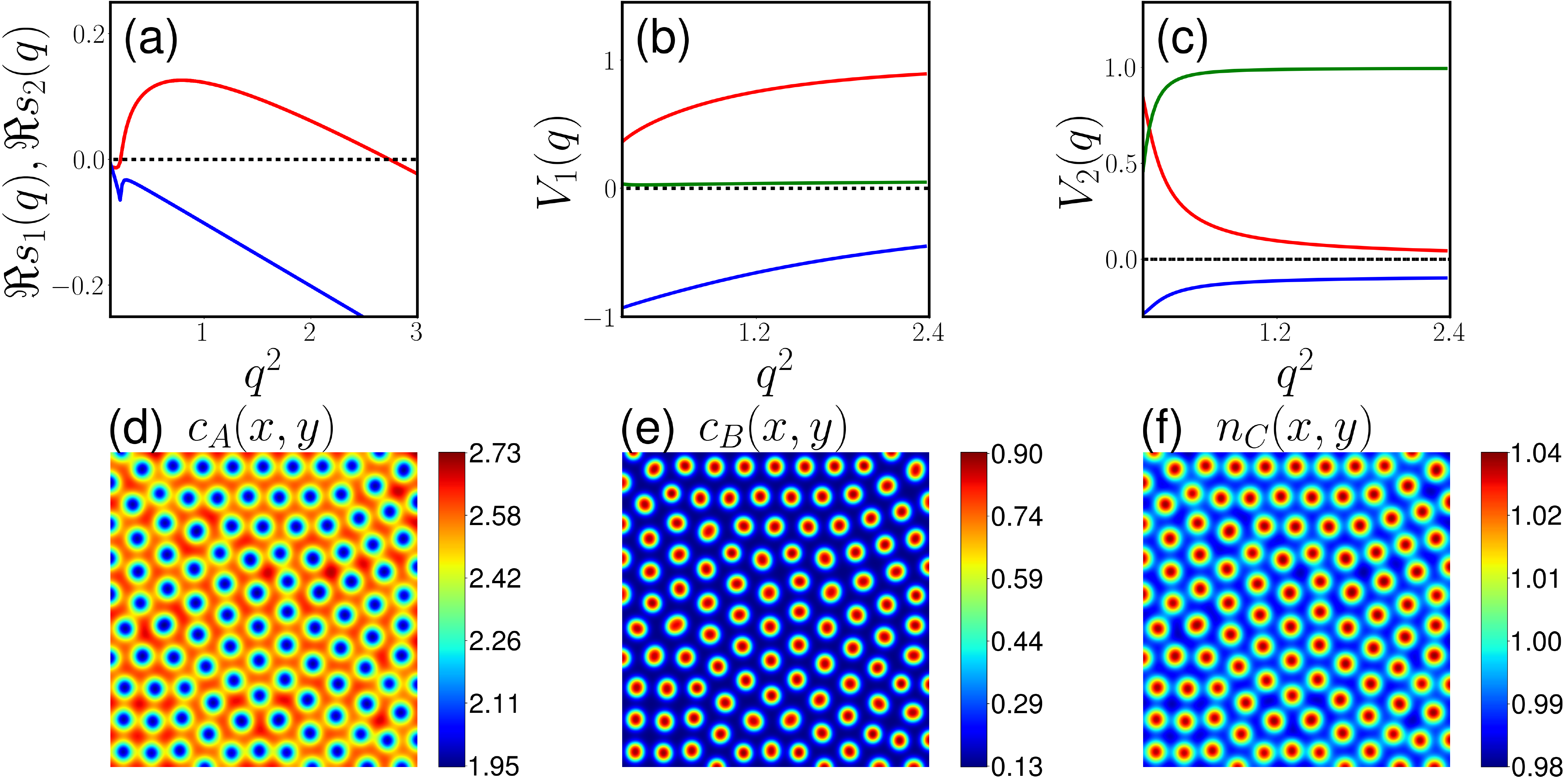}}
\caption{(a) The plot of the real parts of the two largest eigenvalues $\Re s_1(q)$ (red line) and $\Re s_2(q)$ (blue line) versus $q^2$. (b) and (c) Eigenvectors $V_1(q)$ and $V_2(q)$ corresponding to $\Re s_1(q)$ and $\Re s_2(q)$, respectively, versus $q^2$. Here and below, the $\alpha_{iA}(q)$, $\alpha_{iB}(q)$ and $\alpha_{iC}(q)$ coefficients of $V_i(q)$  are denoted by red, blue, and green lines, respectively. (d)-(f) Simulation results for the concentration fields of fuel and product species $c_A(\bm{r})$, $c_B(\bm{r})$ and the motor density field $n_C(\bm{r})$, respectively. The parameters are: $\widetilde{k}_{+} = 0.01$, $\widetilde{D}_{t}^{\rm eff} = 0.1$, $(\mathcal{\widetilde{M}}_{A}, \mathcal{\widetilde{M}}_{B}) = (-0.01, 0.001)$, and $r_{BA} = 0.1$. The elements of $\bm{{\sf R}}$ are $\mathsf{R}_{AA} = -0.194$, $\mathsf{R}_{AB} = -1.315$, $\mathsf{R}_{BA} = 0.094$, $\mathsf{R}_{BB} = 0.315$, and $\mathsf{R}_{AC} = -0.027$.
}
\label{fig:k+01M-_Turing_Pattern}
\end{figure}

When $\widetilde{k}_+ \ne 0$, the motor density is fully coupled to the fluid $c_A$ and $c_B$ concentration fields, and we now explore some of the consequences of this coupling on motor clustering. We take $\widetilde{k}_{+} = 0.01$ and $r_{BA} = 0.1$. Figure~\ref{fig:k+01M-_Turing_Pattern} (a) plots the largest two eigenvalues, $s_1(q)$ and $s_2(q)$, versus $q^2$ for a system with $(\mathcal{\widetilde{M}}_{A}, \mathcal{\widetilde{M}}_{B}) = (-0.01, 0.001)$ and $\widetilde{D}_{t}^{\rm eff} = 0.1$ (other parameters are given in the figure caption), while panels (b) and (c) plot their corresponding eigenvectors, $V_1(q)$ and $V_2(q)$. In this case, there is still a single unstable mode and stationary spot patterns are seen in the concentrations and motor density fields (see movie S2 in the Supplementary Material). In panel (b) we see that the unstable mode $V_1(q)$ has a negligible contribution $\alpha_{1C}$ from the motor density, while for the stable mode $V_2(q)$, $\alpha_{2C}(q)$ is large. Note that spatial variations of the motor density field are not large in this case.

\begin{figure}[htb!]
\centering
\resizebox{1.0\columnwidth}{!}{%
\includegraphics[scale = 1.0]{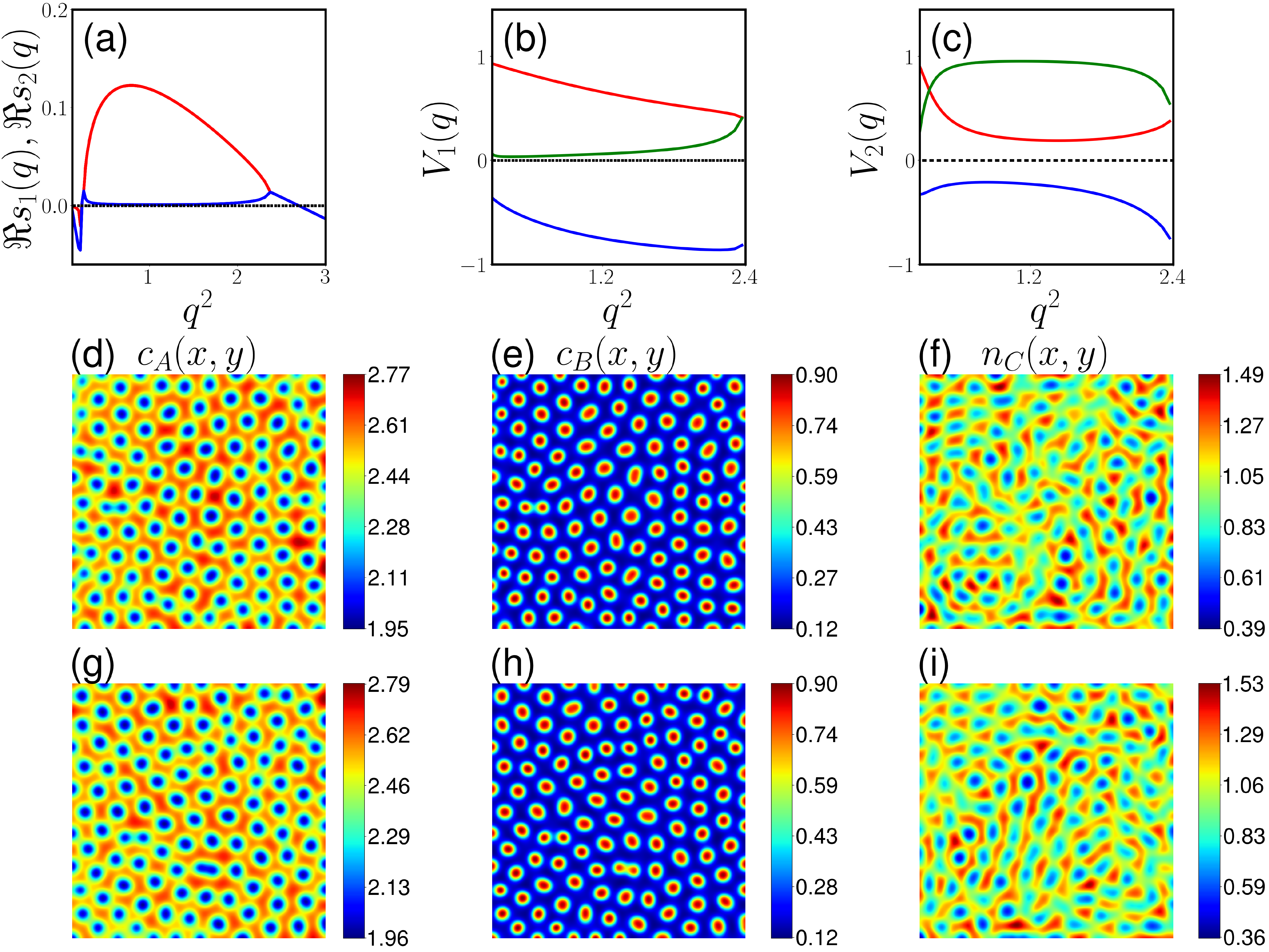}}
\caption{(a) The plot of the real parts of the two largest eigenvalues $\Re s_1(q)$ (red line) and $\Re s_2(q)$ (blue line) versus $q^2$. (b) and (c) Eigenvectors $V_1(q)$ and $V_2(q)$ corresponding to $\Re s_1(q)$ and $\Re s_2(q)$, respectively, versus $q^2$. (d)-(f) Simulation results for the concentration fields of fuel and product species $c_A$, $c_B$ and the motor density field $n_C$, respectively, at time $t = 4000$ and (g)-(i) at $t = 5000$. Parameters are: $\widetilde{k}_{+} = 0.01$, $\widetilde{D}_{t}^{\rm eff} = 0.001$, $(\mathcal{\widetilde{M}}_{A}, \mathcal{\widetilde{M}}_{B}) = (0.01, 0.001)$, and $r_{BA} = 0.1$. The elements of $\bm{{\sf R}}$ are the same as in Fig.~\ref{fig:k+01M-_Turing_Pattern}.}
\label{fig:k+01M+_Turing_Pattern}
\end{figure}

By contrast, as can be seen in Fig.~\ref{fig:k+01M+_Turing_Pattern}, for parameters $\widetilde{k}_{+} = 0.01$, $(\mathcal{\widetilde{M}}_{A}, \mathcal{\widetilde{M}}_{B}) = (0.01, 0.001)$ and $\widetilde{D}_{t}^{\rm eff} = 0.001$ there are two unstable modes, $V_1(q)$ and $V_2(q)$. (The second unstable mode is slightly greater than zero and is difficult to distinguish on the plot.) Now the pattern formation process is very different, as seen in the simulation results in the middle and lower panels for two different times. The spot-like pattern is no longer stationary (see movie S3 in the Supplementary Material). Since the unstable mode $V_2(q)$ has a large $\alpha_{2C}(q)$ coefficient, the motor density field plays a more active role in the dynamic clustering process. Now there are large inhomogeneities in the motor density field, indicating strong dynamic clustering.

\section{Spiral waves} \label{sec:spiral}

Next, we select parameters such that the homogeneous system has an oscillatory solution that arises from a Hopf bifurcation. In the vicinity of such a limit cycle solution, we consider the spatially-distributed system and select the initial condition so that a spiral wave is formed.~\cite{Scheel@SJMA:1998,Kapral_Showalter@Book:1995} We may then study the nature of motor collective motion in this spatially-distributed oscillatory medium. Since the inhomogeneous fluid concentration fields now vary in time, the dynamical response of the active colloid density field to concentration changes depends on its properties and the time scale of the spiral wave oscillations.

\begin{figure}[htb!]
\centering
\resizebox{1.0\columnwidth}{!}{%
\includegraphics[scale = 1.0]{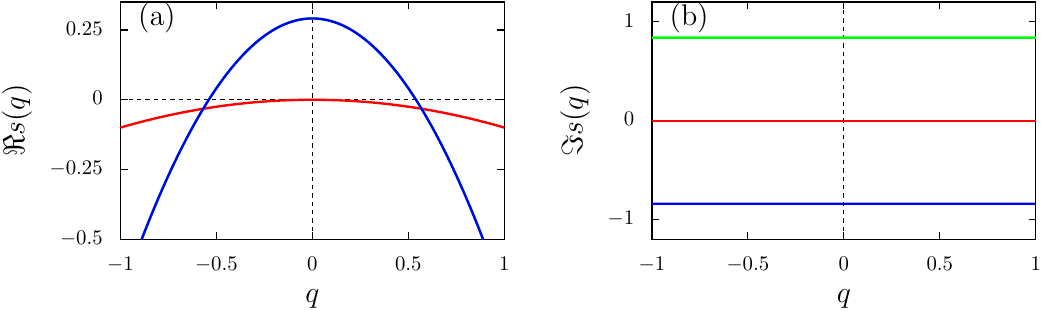}}
\caption{(a) and (b) Plots of the real $\Re s(q)$ and imaginary $\Im s(q)$ parts of the eigenvalues $s_1$, $s_2$ and $s_3$, respectively, versus $q$. The parameters are: $\widetilde{k}_{+} = 0.0025$, $\widetilde{D}_{t}^{\rm eff} = 0.1$, $(\mathcal{\widetilde{M}}_{A}, \mathcal{\widetilde{M}}_{B}) = (-0.01, 0.006)$, and $r_{BA} = 1$. The elements of $\bm{{\sf R}}$ are $\mathsf{R}_{AA} = -0.694$, $\mathsf{R}_{AB} = -3.777$, $\mathsf{R}_{BA} = 0.444$, $\mathsf{R}_{BB} = 1.277$, and $\mathsf{R}_{AC} = -0.009$.}
\label{fig: SA_Spiral_Wave}
\end{figure}

More specifically, we choose the Selkov parameters: $g_{1} = 2.3713,\; g_{2}= 0.25, \; f = 0.25, \; h = 0.4$  and $\widetilde{k}_+ = 0.0025$. For these parameters, we have $\mathsf{R}_{AA} = -0.694$, $\mathsf{R}_{AB} = -3.777$, $\mathsf{R}_{BA} = 0.444$, $\mathsf{R}_{BB} = 1.277$, and $\mathsf{R}_{AC} = -0.009$. The real $\Re s(q)$ and imaginary $\Im s(q)$ parts of the eigenvalues are plotted in Fig.~\ref{fig: SA_Spiral_Wave} as a function of $q$ for parameters $(\mathcal{\widetilde{M}}_{A}, \mathcal{\widetilde{M}}_{B}) = (-0.01, 0.006)$, $\widetilde{D}_{t}^{\rm eff} = 0.1$, and $r_{BA} = 1$. There is a complex conjugate pair of unstable modes within a band of wavenumbers $q$, reflecting the appearance of a spiral wave in the concentrations and motor density fields. (The unstable modes overlap in panel (a), so only one unstable mode is visible.) In particular, for $q = 0.4$, the eigenvectors $V_i(q=0.4)$ corresponding to eigenvalues $s_{1,\; 2} = 0.131 \pm 0.840 \mi$ and $s_3 = -0.016$, respectively, are:
\begin{align}
&
 \begin{pmatrix}
0.946\\
-0.247 - 0.210  \mi \\
-0.001 + 0.002  \mi
\end{pmatrix},
\begin{pmatrix}
0.946\\
-0.247 + 0.210  \mi \\
-0.001 - 0.002  \mi
\end{pmatrix}
\mathrm{and}
\begin{pmatrix}
-0.035\\
0.005\\
0.999
\end{pmatrix},\label{eq:Eigenvectors_Spiral}
\end{align}
where the column entries are ($\alpha_{iA},\alpha_{iB},\alpha_{iC}$). These eigenvectors show that the Selkov kinetics plays a major role in the dynamic clustering process since the first two eigenvectors in Eq.~(\ref{eq:Eigenvectors_Spiral}) corresponding to the unstable $s_{1, \;2}$ eigenvalues have only minimal contributions from the motor density field. Thus, the motor density field adopts a spatial pattern as a response to the gradients in the chemical spiral wave.

\begin{figure}[htb!]
\centering
\resizebox{1.0\columnwidth}{!}{%
\includegraphics[scale = 1.0]{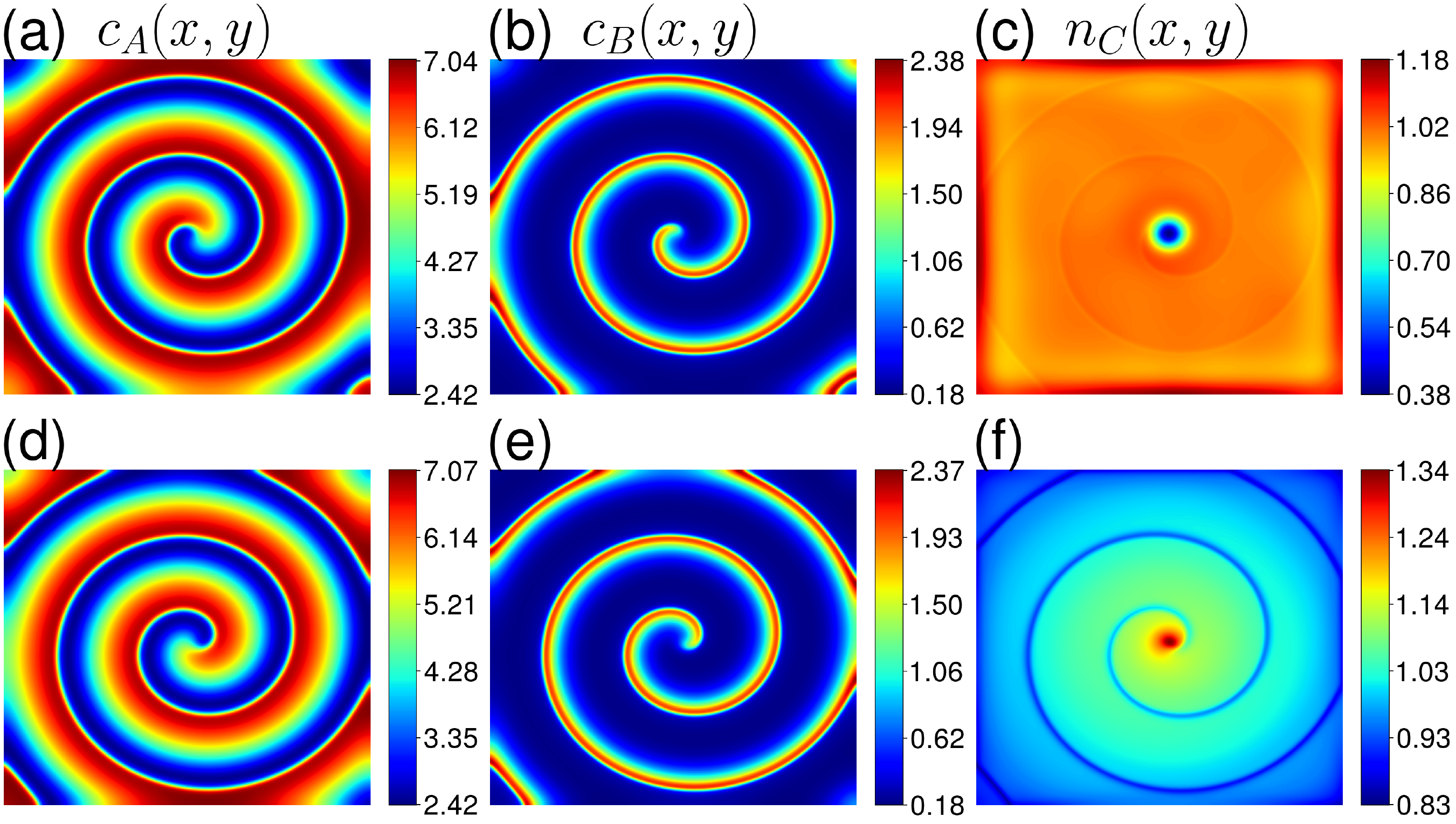}}
\caption{Simulation results for the concentration fields of fuel and product species $c_A$, $c_B$ and the motor density field $n_C$ for $\widetilde{k}_{+} = 0.0025$ and $r_{BA} = 1$ at time $t = 5000$. (a)-(c) Set parameters are: $\widetilde{D}_{t}^{\rm eff} = 0.001$ and $(\mathcal{\widetilde{M}}_{A}, \mathcal{\widetilde{M}}_{B}) = (0.003, 0.006)$, and (d)-(f) for $\widetilde{D}_{t}^{\rm eff} = 0.1$ and $(\mathcal{\widetilde{M}}_{A}, \mathcal{\widetilde{M}}_{B}) = (-0.01, 0.006)$. The elements of $\bm{{\sf R}}$ are the same as in Fig.~\ref{fig: SA_Spiral_Wave}.}
\label{fig: Spiral_Wave_Results}
\end{figure}
Simulation results for the $c_A$, $c_B$ and $n_C$ fields at a time $t = 5000$ are shown in Fig.~\ref{fig: Spiral_Wave_Results}. The chemical concentration fields adopt the form of an outwardly rotating spiral wave, and motor clustering has taken place. For $(\mathcal{\widetilde{M}}_{A}, \mathcal{\widetilde{M}}_{B}) = (0.003, 0.006)$ and $\widetilde{D}_{t}^{\rm eff} = 0.001$, the $n_C$ field is minimum at the center and maximum near the boundaries (see movie S4 in the Supplementary Material). However, for $(\mathcal{\widetilde{M}}_{A}, \mathcal{\widetilde{M}}_{B}) = (-0.01, 0.006)$ and $\widetilde{D}_{t}^{\rm eff} = 0.1$, the $n_C$ field is maximum at the center and decays in a spiral pattern as one moves away from the center of the system (see movie S5 in the Supplementary Material).

Recall that the effective motor diffusion coefficient $\widetilde{D}_{t}^{\rm eff}$ depends on the self-diffusiophoretic velocity $V_{\rm sd}$, while the $\mathcal{\widetilde{M}}_{A,B}$ coefficients depend on    several parameters given earlier, which include $V_{\rm sd}$ as well as the parameters $\xi_k$ that control how the motors respond to gradients. The accumulation or depletion of colloids near the spiral core can be attributed to the different values of $\mathcal{\widetilde{M}}_{A,B}$ in the two cases that determine how the motors respond to chemical gradients. The overall forms of the motor density patterns can be understood from the following considerations: The velocity $V_{\rm sd}$ is much smaller in (a)-(c) than in (d)-(f), with corresponding changes in the value of the effective motor diffusion coefficient. In both cases the spiral wave period is $\tau_{\rm sp} \approx 300$. For $\widetilde{D}_{t}^{\rm eff}=0.001$ the motor will diffuse a distance $\ell_m \approx 0.55$ in a spiral period, while for $\widetilde{D}_{t}^{\rm eff}=0.1$, $\ell_m \approx 5.5$. Therefore, for small $\ell_m$ the motor density field cannot accommodate to the relatively rapidly moving spiral wave and senses an average concentration field; hence, the motor density field in (c) is approximately uniform except near the spiral core. For larger $\ell_m$ in (f) the motor density field can respond to the structure of the spiral concentration gradients, and one sees a depletion of motor density along the spiral arms and a more uniform spiral-shaped density of motors with a maximum at the core. In both (c) and (f) a final spatiotemporal periodic state is established as shown in the figures. Especially in this latter case, motor collective motion is dynamic and takes the form of a rotating spiral wave. In such circumstances, other complex spatiotemporal motor collective states can be formed, such as defect-mediated turbulence,~\cite{coullet1989,Desai_Kapral@Book:2009} if many spiral waves are present in the system.

\section{Discussion}\label{Sec: Conclusions}

The results presented in the text show that the collective behavior of chemically powered motors in chemically active environments that support nonequilibrium patterns differs substantially from that in simple media. In chemically inactive media, the clustering process is governed by various mechanisms mentioned in the Introduction. In particular, we note that chemotactic effects play a major role in the collective dynamics of chemically powered motors~\cite{Huang_et_al@NJP:2017} and can lead to diverse clustering patterns.~\cite{Liebchen_et_al@PRL:2017}

In a chemically active medium that supports waves or other patterns, strong chemical gradients with specific forms exist, and these gradients can have a dominant effect on the diffusiophoretic motor dynamics that causes the motors to accumulate in regions where there are either strong or weak gradients depending on system parameters. We have chosen the Selkov model to illustrate the phenomena since this model has a rich bifurcation structure that includes multiple steady states, Turing bifurcations and spiral waves.~\cite{Richter_et_al@PTP:1981} The Selkov model is especially interesting since the step of glycolysis it describes has been shown to exhibit oscillations and spiral waves in laboratory experiments,~\cite{gly-spiral} and colloids partially coated by enzymes could provide a laboratory realization of some of the results described in this paper. However, other models that support such chemical pattern formation processes could have been used since the chemically active colloids need only participate in the reaction mechanism for the effects to exist. The reactions on the motors need not have the simple linear kinetics used in this work, but could be nonlinear and autocatalytic. Since the reactions on the motors are coupled to those in the fluid phase, the motors can also change the dynamical behavior of chemical patterns in the fluid phase. Instabilities in the motor density equation can drive the entire system into new dynamical regimes, as was seen in the dynamical spot patterns in Fig.~\ref{fig:k+01M+_Turing_Pattern}. While in our studies of spiral wave dynamics system parameters were chosen so that the spiral wave concentration fields control the colloid density fields, in other parameter regimes, especially where spiral wave instabilities exist, coupling to the motor dynamics may also lead to new dynamical system states.

In a more general context, the results show how concentration inhomogeneities that arise in nonequilibrium chemical networks can substantially modify the character of motor collective dynamics. Viewed in the context, the paper provides information that will be useful for potential applications that are likely to involve the ensembles of active agents moving in complex chemical environments.

\section*{Supplementary Material}
Videos showing the spatial and temporal evolution of the chemical and motor patterns shown in the figures in the text are S1 (Fig.~\ref{fig:SA_Turing_Pattern}), S2 (Fig.~\ref{fig:k+01M-_Turing_Pattern}), S3 (Fig.~\ref{fig:k+01M+_Turing_Pattern}) and S4 and S5 (Fig.~\ref{fig: Spiral_Wave_Results}).

\section*{Author Contributions}
Both authors contributed equally to the work.

\section*{Conflicts of interest}
There are no conflicts to declare.

\section*{Data Availability}
The data that supports the findings of this study are available within the article [and its supplementary material].

\section*{Acknowledgements}\label{Sec: Acknowledgment}
This work was supported in part by the Natural Sciences and Engineering Research Council (NSERC) of Canada. Computations were performed on SciNet HPC Consortium computers. SciNet is funded by the Canada Foundation for Innovation, the Government of Ontario, the Ontario Research Excellence Fund, and the University of Toronto.

\appendix
\section{Choice of parameters}\label{Sec: Appendix}

Equation~(\ref{eq:Density_nC}) for the evolution of the local motor density requires as input values of the motor translational and rotational diffusion constants, the motor self-diffusiophoretic velocity $V_{\rm sd}$, as well as the parameters $\zeta_k$, $\xi_k$ and $\lambda_k$ that enter the $\mathcal{M}_{k}$ coefficients. The latter three parameters and $V_{\rm sd}$ depend on the reaction rate $k_+$.~\cite{Gaspard_Kapral@Research:2020}

To estimate these parameters, we use typical values for a system of micromotors: radius $R \sim 1 ~{\rm \mu m}$, $D_{t} \sim 10^{-13} ~{\rm m^{2}/s}$, $D_{r} \sim 0.2 ~{\rm s^{-1}}$, $V_{\rm sd} \sim 1-10 ~{\rm \mu m/s}$, radii of fuel $A$ and product $B$ species $a \sim 1 ~{\rm nm}$, $D_{A} \sim 10^{-9} ~{\rm m^{2}/s}$, $D_{B} \sim 10^{-10}-10^{-9} ~{\rm m^{2}/s}$ and diffusiophoretic coefficients $\lvert b_{k} \rvert \sim 10^{-36}- 6 \times 10^{-36} ~{\rm m^{5}/s}$, that depend on the temperature, viscosity and interaction potentials.

The density of motors is approximately $10^{-8}- 10^{-7} ~{\rm mol/m^{3}}$ corresponding to a volume fraction $\phi \sim 0.02-0.2$. The reaction rate is assumed to be diffusion-limited with value $k_{+} \sim 10^{-14} ~{\rm m^{3}/s}$. Correspondingly, the reaction rate per unit surface area of the motor is calculated as $\kappa_{+} \sim 10^{-3} ~{\rm m/s}$. The volume of the quasi-two-dimensional square is $V \sim 10 \times 10 \times 0.1~{\rm mm^{3}}$, the temperature of the system is $T \sim 300~{\rm K}$ and the viscosity of the solution is $\mu \sim 10^{-3} ~{\rm kg/(m~s)}$. These values were used to compute the dimensionless parameters in the text.



\bibliography{rsc} 

\begin{thebibliography}{60}
\expandafter\ifx\csname natexlab\endcsname\relax\def\natexlab#1{#1}\fi
\expandafter\ifx\csname bibnamefont\endcsname\relax
  \def\bibnamefont#1{#1}\fi
\expandafter\ifx\csname bibfnamefont\endcsname\relax
  \def\bibfnamefont#1{#1}\fi
\expandafter\ifx\csname citenamefont\endcsname\relax
  \def\citenamefont#1{#1}\fi
\expandafter\ifx\csname url\endcsname\relax
  \def\url#1{\texttt{#1}}\fi
\expandafter\ifx\csname urlprefix\endcsname\relax\def\urlprefix{URL }\fi
\providecommand{\bibinfo}[2]{#2}
\providecommand{\eprint}[2][]{\url{#2}}

\bibitem[{\citenamefont{Gompper et~al.}(2020)}]{roadmap2020}
\bibinfo{author}{\bibfnamefont{G.}~\bibnamefont{Gompper}} \bibnamefont{et~al.},
  \bibinfo{journal}{J. Phys.: Condens. Matter} \textbf{\bibinfo{volume}{32}},
  \bibinfo{pages}{193001} (\bibinfo{year}{2020}).

\bibitem[{\citenamefont{Duan et~al.}(2015)\citenamefont{Duan, Wang, Das, Yadav,
  Mallouk, and Sen}}]{DWDYMS15}
\bibinfo{author}{\bibfnamefont{W.}~\bibnamefont{Duan}},
  \bibinfo{author}{\bibfnamefont{W.}~\bibnamefont{Wang}},
  \bibinfo{author}{\bibfnamefont{S.}~\bibnamefont{Das}},
  \bibinfo{author}{\bibfnamefont{V.}~\bibnamefont{Yadav}},
  \bibinfo{author}{\bibfnamefont{T.~E.} \bibnamefont{Mallouk}},
  \bibnamefont{and} \bibinfo{author}{\bibfnamefont{A.}~\bibnamefont{Sen}},
  \bibinfo{journal}{Annu. Rev. Anal. Chem.} \textbf{\bibinfo{volume}{8}},
  \bibinfo{pages}{311} (\bibinfo{year}{2015}).

\bibitem[{\citenamefont{Ma et~al.}(2015{\natexlab{a}})\citenamefont{Ma, Hahn,
  and S{\'a}nchez}}]{MHS15}
\bibinfo{author}{\bibfnamefont{X.}~\bibnamefont{Ma}},
  \bibinfo{author}{\bibfnamefont{K.}~\bibnamefont{Hahn}}, \bibnamefont{and}
  \bibinfo{author}{\bibfnamefont{S.}~\bibnamefont{S{\'a}nchez}},
  \bibinfo{journal}{J. Am. Chem. Soc.} \textbf{\bibinfo{volume}{137}},
  \bibinfo{pages}{4976} (\bibinfo{year}{2015}{\natexlab{a}}).

\bibitem[{\citenamefont{Wang et~al.}(2018)\citenamefont{Wang, Xiong, Zheng,
  Zhan, and Tang}}]{Jizhuang2018}
\bibinfo{author}{\bibfnamefont{J.}~\bibnamefont{Wang}},
  \bibinfo{author}{\bibfnamefont{Z.}~\bibnamefont{Xiong}},
  \bibinfo{author}{\bibfnamefont{J.}~\bibnamefont{Zheng}},
  \bibinfo{author}{\bibfnamefont{X.}~\bibnamefont{Zhan}}, \bibnamefont{and}
  \bibinfo{author}{\bibfnamefont{J.}~\bibnamefont{Tang}},
  \bibinfo{journal}{Acc. Chem. Res.} \textbf{\bibinfo{volume}{51}},
  \bibinfo{pages}{1957} (\bibinfo{year}{2018}).

\bibitem[{\citenamefont{Wang et~al.}(2019)\citenamefont{Wang, Liu, Wang, Peng,
  and Tu}}]{wang2019application}
\bibinfo{author}{\bibfnamefont{S.}~\bibnamefont{Wang}},
  \bibinfo{author}{\bibfnamefont{K.}~\bibnamefont{Liu}},
  \bibinfo{author}{\bibfnamefont{F.}~\bibnamefont{Wang}},
  \bibinfo{author}{\bibfnamefont{F.}~\bibnamefont{Peng}}, \bibnamefont{and}
  \bibinfo{author}{\bibfnamefont{Y.}~\bibnamefont{Tu}}, \bibinfo{journal}{Chem.
  Asian J.} \textbf{\bibinfo{volume}{14}}, \bibinfo{pages}{2336}
  (\bibinfo{year}{2019}).

\bibitem[{\citenamefont{Xu et~al.}(2019)\citenamefont{Xu, Wang, Liang, You,
  Sanchez, and Ma}}]{xu2019self}
\bibinfo{author}{\bibfnamefont{D.}~\bibnamefont{Xu}},
  \bibinfo{author}{\bibfnamefont{Y.}~\bibnamefont{Wang}},
  \bibinfo{author}{\bibfnamefont{C.}~\bibnamefont{Liang}},
  \bibinfo{author}{\bibfnamefont{Y.}~\bibnamefont{You}},
  \bibinfo{author}{\bibfnamefont{S.}~\bibnamefont{Sanchez}}, \bibnamefont{and}
  \bibinfo{author}{\bibfnamefont{X.}~\bibnamefont{Ma}},
  \bibinfo{journal}{Small} \textbf{\bibinfo{volume}{16}},
  \bibinfo{pages}{1902464} (\bibinfo{year}{2019}).

\bibitem[{\citenamefont{Ying and Pumera}(2019)}]{ying2019micro}
\bibinfo{author}{\bibfnamefont{Y.}~\bibnamefont{Ying}} \bibnamefont{and}
  \bibinfo{author}{\bibfnamefont{M.}~\bibnamefont{Pumera}},
  \bibinfo{journal}{Chem. Eur. J.} \textbf{\bibinfo{volume}{25}},
  \bibinfo{pages}{106} (\bibinfo{year}{2019}).

\bibitem[{\citenamefont{Soto et~al.}(2020)\citenamefont{Soto, Kupor,
  Lopez-Ramirez, Wei, Karshalev, Tang, Tehrani, and Wang}}]{soto2019onion}
\bibinfo{author}{\bibfnamefont{F.}~\bibnamefont{Soto}},
  \bibinfo{author}{\bibfnamefont{D.}~\bibnamefont{Kupor}},
  \bibinfo{author}{\bibfnamefont{M.~A.} \bibnamefont{Lopez-Ramirez}},
  \bibinfo{author}{\bibfnamefont{F.}~\bibnamefont{Wei}},
  \bibinfo{author}{\bibfnamefont{E.}~\bibnamefont{Karshalev}},
  \bibinfo{author}{\bibfnamefont{S.}~\bibnamefont{Tang}},
  \bibinfo{author}{\bibfnamefont{F.}~\bibnamefont{Tehrani}}, \bibnamefont{and}
  \bibinfo{author}{\bibfnamefont{J.}~\bibnamefont{Wang}},
  \bibinfo{journal}{Angew. Chem. Int. Ed.} \textbf{\bibinfo{volume}{59}},
  \bibinfo{pages}{3480} (\bibinfo{year}{2020}).

\bibitem[{\citenamefont{Tang et~al.}(2020)}]{Tang20}
\bibinfo{author}{\bibfnamefont{S.}~\bibnamefont{Tang}} \bibnamefont{et~al.},
  \bibinfo{journal}{Sci. Robot.} \textbf{\bibinfo{volume}{5}},
  \bibinfo{pages}{eaba6137} (\bibinfo{year}{2020}).

\bibitem[{\citenamefont{Wang}(2013)}]{wangbook2013}
\bibinfo{author}{\bibfnamefont{J.}~\bibnamefont{Wang}},
  \emph{\bibinfo{title}{Nanomachines: Fundamentals and applications}}
  (\bibinfo{publisher}{Wiley-VCH}, \bibinfo{address}{Weinheim},
  \bibinfo{year}{2013}).

\bibitem[{\citenamefont{Anderson}(1983)}]{anderson1983}
\bibinfo{author}{\bibfnamefont{J.~L.} \bibnamefont{Anderson}},
  \bibinfo{journal}{Phys. Fluids} \textbf{\bibinfo{volume}{26}},
  \bibinfo{pages}{2871} (\bibinfo{year}{1983}).

\bibitem[{\citenamefont{Golestanian et~al.}(2007)\citenamefont{Golestanian,
  Liverpool, and Ajdari}}]{golestanian2007}
\bibinfo{author}{\bibfnamefont{R.}~\bibnamefont{Golestanian}},
  \bibinfo{author}{\bibfnamefont{T.~B.} \bibnamefont{Liverpool}},
  \bibnamefont{and} \bibinfo{author}{\bibfnamefont{A.}~\bibnamefont{Ajdari}},
  \bibinfo{journal}{New J. Phys.} \textbf{\bibinfo{volume}{9}},
  \bibinfo{pages}{126} (\bibinfo{year}{2007}).

\bibitem[{\citenamefont{Kapral}(2013)}]{Kapral@JCP:2013}
\bibinfo{author}{\bibfnamefont{R.}~\bibnamefont{Kapral}}, \bibinfo{journal}{J.
  Chem. Phys.} \textbf{\bibinfo{volume}{138}}, \bibinfo{pages}{020901}
  (\bibinfo{year}{2013}).

\bibitem[{\citenamefont{Wang et~al.}(2013)\citenamefont{Wang, Duan, Ahmed,
  Mallouk, and Sen}}]{SenRev2013}
\bibinfo{author}{\bibfnamefont{W.}~\bibnamefont{Wang}},
  \bibinfo{author}{\bibfnamefont{W.}~\bibnamefont{Duan}},
  \bibinfo{author}{\bibfnamefont{S.}~\bibnamefont{Ahmed}},
  \bibinfo{author}{\bibfnamefont{T.~E.} \bibnamefont{Mallouk}},
  \bibnamefont{and} \bibinfo{author}{\bibfnamefont{A.}~\bibnamefont{Sen}},
  \bibinfo{journal}{Nano Today} \textbf{\bibinfo{volume}{8}},
  \bibinfo{pages}{531} (\bibinfo{year}{2013}).

\bibitem[{\citenamefont{Colberg et~al.}(2014)\citenamefont{Colberg, Reigh,
  Robertson, and Kapral}}]{colberg14}
\bibinfo{author}{\bibfnamefont{P.~H.} \bibnamefont{Colberg}},
  \bibinfo{author}{\bibfnamefont{S.~Y.} \bibnamefont{Reigh}},
  \bibinfo{author}{\bibfnamefont{B.}~\bibnamefont{Robertson}},
  \bibnamefont{and} \bibinfo{author}{\bibfnamefont{R.}~\bibnamefont{Kapral}},
  \bibinfo{journal}{Acc. Chem. Res.} \textbf{\bibinfo{volume}{47}},
  \bibinfo{pages}{3504} (\bibinfo{year}{2014}).

\bibitem[{\citenamefont{S{\'a}nchez et~al.}(2015)\citenamefont{S{\'a}nchez,
  Soler, and Katuri}}]{sanchez14}
\bibinfo{author}{\bibfnamefont{S.}~\bibnamefont{S{\'a}nchez}},
  \bibinfo{author}{\bibfnamefont{L.}~\bibnamefont{Soler}}, \bibnamefont{and}
  \bibinfo{author}{\bibfnamefont{J.}~\bibnamefont{Katuri}},
  \bibinfo{journal}{Angew. Chem. Int. Ed.} \textbf{\bibinfo{volume}{54}},
  \bibinfo{pages}{1414} (\bibinfo{year}{2015}).

\bibitem[{\citenamefont{Yadav et~al.}(2015)\citenamefont{Yadav, Duan, Butler,
  and Sen}}]{Yadav2015}
\bibinfo{author}{\bibfnamefont{V.}~\bibnamefont{Yadav}},
  \bibinfo{author}{\bibfnamefont{W.}~\bibnamefont{Duan}},
  \bibinfo{author}{\bibfnamefont{P.~J.} \bibnamefont{Butler}},
  \bibnamefont{and} \bibinfo{author}{\bibfnamefont{A.}~\bibnamefont{Sen}},
  \bibinfo{journal}{Annual Review of Biophysics} \textbf{\bibinfo{volume}{44}},
  \bibinfo{pages}{77} (\bibinfo{year}{2015}).

\bibitem[{\citenamefont{Oshanin et~al.}(2017)\citenamefont{Oshanin, Popescu,
  and Dietrich}}]{OPD17}
\bibinfo{author}{\bibfnamefont{G.}~\bibnamefont{Oshanin}},
  \bibinfo{author}{\bibfnamefont{M.~N.} \bibnamefont{Popescu}},
  \bibnamefont{and} \bibinfo{author}{\bibfnamefont{S.}~\bibnamefont{Dietrich}},
  \bibinfo{journal}{J. Phys. A: Math. Theor.} \textbf{\bibinfo{volume}{50}},
  \bibinfo{pages}{134001} (\bibinfo{year}{2017}).

\bibitem[{\citenamefont{Gaspard and Kapral}(2019)}]{GK18a}
\bibinfo{author}{\bibfnamefont{P.}~\bibnamefont{Gaspard}} \bibnamefont{and}
  \bibinfo{author}{\bibfnamefont{R.}~\bibnamefont{Kapral}},
  \bibinfo{journal}{Adv. Phys. X} \textbf{\bibinfo{volume}{4}},
  \bibinfo{pages}{1602480} (\bibinfo{year}{2019}).

\bibitem[{\citenamefont{Howse et~al.}(2007)\citenamefont{Howse, Jones, Ryan,
  Gough, Vafabakhsh, and Golestanian}}]{Howse_et_al@PRL:2007}
\bibinfo{author}{\bibfnamefont{J.~R.} \bibnamefont{Howse}},
  \bibinfo{author}{\bibfnamefont{R.~A.~L.} \bibnamefont{Jones}},
  \bibinfo{author}{\bibfnamefont{A.~J.} \bibnamefont{Ryan}},
  \bibinfo{author}{\bibfnamefont{T.}~\bibnamefont{Gough}},
  \bibinfo{author}{\bibfnamefont{R.}~\bibnamefont{Vafabakhsh}},
  \bibnamefont{and}
  \bibinfo{author}{\bibfnamefont{R.}~\bibnamefont{Golestanian}},
  \bibinfo{journal}{Phys. Rev. Lett.} \textbf{\bibinfo{volume}{99}},
  \bibinfo{pages}{048102} (\bibinfo{year}{2007}).

\bibitem[{\citenamefont{R\"uckner and Kapral}(2007)}]{Ruckner_Kapral@PRL:2007}
\bibinfo{author}{\bibfnamefont{G.}~\bibnamefont{R\"uckner}} \bibnamefont{and}
  \bibinfo{author}{\bibfnamefont{R.}~\bibnamefont{Kapral}},
  \bibinfo{journal}{Phys. Rev. Lett.} \textbf{\bibinfo{volume}{98}},
  \bibinfo{pages}{150603} (\bibinfo{year}{2007}).

\bibitem[{\citenamefont{de~Buyl and Kapral}(2013)}]{dBK13}
\bibinfo{author}{\bibfnamefont{P.}~\bibnamefont{de~Buyl}} \bibnamefont{and}
  \bibinfo{author}{\bibfnamefont{R.}~\bibnamefont{Kapral}},
  \bibinfo{journal}{Nanoscale} \textbf{\bibinfo{volume}{5}},
  \bibinfo{pages}{1337} (\bibinfo{year}{2013}).

\bibitem[{\citenamefont{Dey et~al.}(2015)\citenamefont{Dey, Zhao, Tansi,
  M{\'e}ndez-Ortiz, C{\'o}rdova-Figueroa, Golestanian, and Sen}}]{Dey2015}
\bibinfo{author}{\bibfnamefont{K.~K.} \bibnamefont{Dey}},
  \bibinfo{author}{\bibfnamefont{X.}~\bibnamefont{Zhao}},
  \bibinfo{author}{\bibfnamefont{B.~M.} \bibnamefont{Tansi}},
  \bibinfo{author}{\bibfnamefont{W.~J.} \bibnamefont{M{\'e}ndez-Ortiz}},
  \bibinfo{author}{\bibfnamefont{U.~M.} \bibnamefont{C{\'o}rdova-Figueroa}},
  \bibinfo{author}{\bibfnamefont{R.}~\bibnamefont{Golestanian}},
  \bibnamefont{and} \bibinfo{author}{\bibfnamefont{A.}~\bibnamefont{Sen}},
  \bibinfo{journal}{Nano Lett.} \textbf{\bibinfo{volume}{15}},
  \bibinfo{pages}{8311} (\bibinfo{year}{2015}).

\bibitem[{\citenamefont{Ma et~al.}(2015{\natexlab{b}})\citenamefont{Ma,
  Jannasch, Albrecht, Hahn, Miguel-Lopez, Schaffer, and
  S{\'a}nchez}}]{MJAHMSS15}
\bibinfo{author}{\bibfnamefont{X.}~\bibnamefont{Ma}},
  \bibinfo{author}{\bibfnamefont{A.}~\bibnamefont{Jannasch}},
  \bibinfo{author}{\bibfnamefont{U.~R.} \bibnamefont{Albrecht}},
  \bibinfo{author}{\bibfnamefont{K.}~\bibnamefont{Hahn}},
  \bibinfo{author}{\bibfnamefont{A.}~\bibnamefont{Miguel-Lopez}},
  \bibinfo{author}{\bibfnamefont{E.}~\bibnamefont{Schaffer}}, \bibnamefont{and}
  \bibinfo{author}{\bibfnamefont{S.}~\bibnamefont{S{\'a}nchez}},
  \bibinfo{journal}{Nano Lett.} \textbf{\bibinfo{volume}{15}},
  \bibinfo{pages}{7043} (\bibinfo{year}{2015}{\natexlab{b}}).

\bibitem[{\citenamefont{Ma et~al.}(2016)\citenamefont{Ma, Hortelao, Patino, and
  S{\'a}nchez}}]{MHPS16}
\bibinfo{author}{\bibfnamefont{X.}~\bibnamefont{Ma}},
  \bibinfo{author}{\bibfnamefont{A.~C.} \bibnamefont{Hortelao}},
  \bibinfo{author}{\bibfnamefont{T.}~\bibnamefont{Patino}}, \bibnamefont{and}
  \bibinfo{author}{\bibfnamefont{S.}~\bibnamefont{S{\'a}nchez}},
  \bibinfo{journal}{ACS Nano} \textbf{\bibinfo{volume}{10}},
  \bibinfo{pages}{9111} (\bibinfo{year}{2016}).

\bibitem[{\citenamefont{Zhao et~al.}(2018)\citenamefont{Zhao, Gentile,
  Mohajerani, and Sen}}]{ZGMS18}
\bibinfo{author}{\bibfnamefont{X.}~\bibnamefont{Zhao}},
  \bibinfo{author}{\bibfnamefont{K.}~\bibnamefont{Gentile}},
  \bibinfo{author}{\bibfnamefont{F.}~\bibnamefont{Mohajerani}},
  \bibnamefont{and} \bibinfo{author}{\bibfnamefont{A.}~\bibnamefont{Sen}},
  \bibinfo{journal}{Acc. Chem. Res.} \textbf{\bibinfo{volume}{51}},
  \bibinfo{pages}{2373} (\bibinfo{year}{2018}).

\bibitem[{\citenamefont{Theurkauff et~al.}(2012)\citenamefont{Theurkauff,
  Cottin-Bizonne, Palacci, Ybert, and Bocquet}}]{Theurkauff_et_al@PRL:2012}
\bibinfo{author}{\bibfnamefont{I.}~\bibnamefont{Theurkauff}},
  \bibinfo{author}{\bibfnamefont{C.}~\bibnamefont{Cottin-Bizonne}},
  \bibinfo{author}{\bibfnamefont{J.}~\bibnamefont{Palacci}},
  \bibinfo{author}{\bibfnamefont{C.}~\bibnamefont{Ybert}}, \bibnamefont{and}
  \bibinfo{author}{\bibfnamefont{L.}~\bibnamefont{Bocquet}},
  \bibinfo{journal}{Phys. Rev. Lett.} \textbf{\bibinfo{volume}{108}},
  \bibinfo{pages}{268303} (\bibinfo{year}{2012}).

\bibitem[{\citenamefont{Thakur and Kapral}(2012)}]{Thakur_Kapral@PRE:2012}
\bibinfo{author}{\bibfnamefont{S.}~\bibnamefont{Thakur}} \bibnamefont{and}
  \bibinfo{author}{\bibfnamefont{R.}~\bibnamefont{Kapral}},
  \bibinfo{journal}{Phys. Rev. E} \textbf{\bibinfo{volume}{85}},
  \bibinfo{pages}{026121} (\bibinfo{year}{2012}).

\bibitem[{\citenamefont{Gao et~al.}(2014)\citenamefont{Gao, Pei, Dong, and
  Wang}}]{Gao_et_al@JACS:2014}
\bibinfo{author}{\bibfnamefont{W.}~\bibnamefont{Gao}},
  \bibinfo{author}{\bibfnamefont{A.}~\bibnamefont{Pei}},
  \bibinfo{author}{\bibfnamefont{R.}~\bibnamefont{Dong}}, \bibnamefont{and}
  \bibinfo{author}{\bibfnamefont{J.}~\bibnamefont{Wang}}, \bibinfo{journal}{J.
  Am. Chem. Soc.} \textbf{\bibinfo{volume}{136}}, \bibinfo{pages}{2276}
  (\bibinfo{year}{2014}).

\bibitem[{\citenamefont{Cates and Tailleur}(2015)}]{Cates_Tailleur@ARCMP:2015}
\bibinfo{author}{\bibfnamefont{M.~E.} \bibnamefont{Cates}} \bibnamefont{and}
  \bibinfo{author}{\bibfnamefont{J.}~\bibnamefont{Tailleur}},
  \bibinfo{journal}{Annu. Rev. Condens. Matter Phys.}
  \textbf{\bibinfo{volume}{6}}, \bibinfo{pages}{219} (\bibinfo{year}{2015}).

\bibitem[{\citenamefont{Saha et~al.}(2014)\citenamefont{Saha, Golestanian, and
  Ramaswamy}}]{Saha_et_al@PRE:2014}
\bibinfo{author}{\bibfnamefont{S.}~\bibnamefont{Saha}},
  \bibinfo{author}{\bibfnamefont{R.}~\bibnamefont{Golestanian}},
  \bibnamefont{and}
  \bibinfo{author}{\bibfnamefont{S.}~\bibnamefont{Ramaswamy}},
  \bibinfo{journal}{Phys. Rev. E} \textbf{\bibinfo{volume}{89}},
  \bibinfo{pages}{062316} (\bibinfo{year}{2014}).

\bibitem[{\citenamefont{Wang et~al.}(2015)\citenamefont{Wang, Duan, Ahmed, Sen,
  and Mallouk}}]{Wang_et_al@ACR:2015}
\bibinfo{author}{\bibfnamefont{W.}~\bibnamefont{Wang}},
  \bibinfo{author}{\bibfnamefont{W.}~\bibnamefont{Duan}},
  \bibinfo{author}{\bibfnamefont{S.}~\bibnamefont{Ahmed}},
  \bibinfo{author}{\bibfnamefont{A.}~\bibnamefont{Sen}}, \bibnamefont{and}
  \bibinfo{author}{\bibfnamefont{T.~E.} \bibnamefont{Mallouk}},
  \bibinfo{journal}{Acc. Chem. Res.} \textbf{\bibinfo{volume}{48}},
  \bibinfo{pages}{1938} (\bibinfo{year}{2015}).

\bibitem[{\citenamefont{Liebchen et~al.}(2017)\citenamefont{Liebchen,
  Marenduzzo, and Cates}}]{Liebchen_et_al@PRL:2017}
\bibinfo{author}{\bibfnamefont{B.}~\bibnamefont{Liebchen}},
  \bibinfo{author}{\bibfnamefont{D.}~\bibnamefont{Marenduzzo}},
  \bibnamefont{and} \bibinfo{author}{\bibfnamefont{M.~E.} \bibnamefont{Cates}},
  \bibinfo{journal}{Phys. Rev. Lett.} \textbf{\bibinfo{volume}{118}},
  \bibinfo{pages}{268001} (\bibinfo{year}{2017}).

\bibitem[{\citenamefont{Huang et~al.}(2017)\citenamefont{Huang, Schofield, and
  Kapral}}]{Huang_et_al@NJP:2017}
\bibinfo{author}{\bibfnamefont{M.~J.} \bibnamefont{Huang}},
  \bibinfo{author}{\bibfnamefont{J.}~\bibnamefont{Schofield}},
  \bibnamefont{and} \bibinfo{author}{\bibfnamefont{R.}~\bibnamefont{Kapral}},
  \bibinfo{journal}{New J. Phys.} \textbf{\bibinfo{volume}{19}},
  \bibinfo{pages}{125003} (\bibinfo{year}{2017}).

\bibitem[{\citenamefont{Illien et~al.}(2017)\citenamefont{Illien, Golestanian,
  and Sen}}]{Illien2017}
\bibinfo{author}{\bibfnamefont{P.}~\bibnamefont{Illien}},
  \bibinfo{author}{\bibfnamefont{R.}~\bibnamefont{Golestanian}},
  \bibnamefont{and} \bibinfo{author}{\bibfnamefont{A.}~\bibnamefont{Sen}},
  \bibinfo{journal}{Chem. Soc. Rev.} \textbf{\bibinfo{volume}{46}},
  \bibinfo{pages}{5508} (\bibinfo{year}{2017}).

\bibitem[{\citenamefont{Ginot et~al.}(2018)\citenamefont{Ginot, Theurkauff,
  Detcheverry, Ybert, and Cottin-Bizonne}}]{Ginot_et_al@NC:2018}
\bibinfo{author}{\bibfnamefont{F.}~\bibnamefont{Ginot}},
  \bibinfo{author}{\bibfnamefont{I.}~\bibnamefont{Theurkauff}},
  \bibinfo{author}{\bibfnamefont{F.}~\bibnamefont{Detcheverry}},
  \bibinfo{author}{\bibfnamefont{C.}~\bibnamefont{Ybert}}, \bibnamefont{and}
  \bibinfo{author}{\bibfnamefont{C.}~\bibnamefont{Cottin-Bizonne}},
  \bibinfo{journal}{Nature Communications} \textbf{\bibinfo{volume}{9}},
  \bibinfo{pages}{696} (\bibinfo{year}{2018}).

\bibitem[{\citenamefont{Pohl and Stark}(2015)}]{Pohl_Stark@NC:2018}
\bibinfo{author}{\bibfnamefont{O.}~\bibnamefont{Pohl}} \bibnamefont{and}
  \bibinfo{author}{\bibfnamefont{H.}~\bibnamefont{Stark}},
  \bibinfo{journal}{Eur. Phys. J. E} \textbf{\bibinfo{volume}{38}},
  \bibinfo{pages}{93} (\bibinfo{year}{2015}).

\bibitem[{\citenamefont{Huang et~al.}(2019)\citenamefont{Huang, Schofield,
  Gaspard, and Kapral}}]{Huang_et_al@JCP:2019}
\bibinfo{author}{\bibfnamefont{M.~J.} \bibnamefont{Huang}},
  \bibinfo{author}{\bibfnamefont{J.}~\bibnamefont{Schofield}},
  \bibinfo{author}{\bibfnamefont{P.}~\bibnamefont{Gaspard}}, \bibnamefont{and}
  \bibinfo{author}{\bibfnamefont{R.}~\bibnamefont{Kapral}},
  \bibinfo{journal}{J. Chem. Phys.} \textbf{\bibinfo{volume}{150}},
  \bibinfo{pages}{124110} (\bibinfo{year}{2019}).

\bibitem[{\citenamefont{Stark}(2018)}]{Stark@ACR:2018}
\bibinfo{author}{\bibfnamefont{H.}~\bibnamefont{Stark}}, \bibinfo{journal}{Acc.
  Chem. Res.} \textbf{\bibinfo{volume}{51}}, \bibinfo{pages}{2681}
  (\bibinfo{year}{2018}).

\bibitem[{\citenamefont{Robertson et~al.}(2018)\citenamefont{Robertson, Huang,
  Chen, and Kapral}}]{CRRK18}
\bibinfo{author}{\bibfnamefont{B.}~\bibnamefont{Robertson}},
  \bibinfo{author}{\bibfnamefont{M.~J.} \bibnamefont{Huang}},
  \bibinfo{author}{\bibfnamefont{J.~X.} \bibnamefont{Chen}}, \bibnamefont{and}
  \bibinfo{author}{\bibfnamefont{R.}~\bibnamefont{Kapral}},
  \bibinfo{journal}{Acc. Chem. Res.} \textbf{\bibinfo{volume}{51}},
  \bibinfo{pages}{2355} (\bibinfo{year}{2018}).

\bibitem[{\citenamefont{Bechinger et~al.}(2016)\citenamefont{Bechinger,
  Leonardo, L\"owen, Reichhardt, Volpe, and Volpe}}]{Bechinger_et_al@RMP:2016}
\bibinfo{author}{\bibfnamefont{C.}~\bibnamefont{Bechinger}},
  \bibinfo{author}{\bibfnamefont{R.~D.} \bibnamefont{Leonardo}},
  \bibinfo{author}{\bibfnamefont{H.}~\bibnamefont{L\"owen}},
  \bibinfo{author}{\bibfnamefont{C.}~\bibnamefont{Reichhardt}},
  \bibinfo{author}{\bibfnamefont{G.}~\bibnamefont{Volpe}}, \bibnamefont{and}
  \bibinfo{author}{\bibfnamefont{G.}~\bibnamefont{Volpe}},
  \bibinfo{journal}{Rev. Mod. Phys.} \textbf{\bibinfo{volume}{88}},
  \bibinfo{pages}{045006} (\bibinfo{year}{2016}).

\bibitem[{\citenamefont{Thakur and Kapral}(2011)}]{thakur:11_2}
\bibinfo{author}{\bibfnamefont{S.}~\bibnamefont{Thakur}} \bibnamefont{and}
  \bibinfo{author}{\bibfnamefont{R.}~\bibnamefont{Kapral}},
  \bibinfo{journal}{J. Chem. Phys.} \textbf{\bibinfo{volume}{135}},
  \bibinfo{pages}{024509} (\bibinfo{year}{2011}).

\bibitem[{\citenamefont{Robertson and
  Kapral}(2015)}]{Robertson_Kapral@JCP:2015}
\bibinfo{author}{\bibfnamefont{B.}~\bibnamefont{Robertson}} \bibnamefont{and}
  \bibinfo{author}{\bibfnamefont{R.}~\bibnamefont{Kapral}},
  \bibinfo{journal}{J. Chem. Phys.} \textbf{\bibinfo{volume}{142}},
  \bibinfo{pages}{154902} (\bibinfo{year}{2015}).

\bibitem[{\citenamefont{Thakur et~al.}(2011)\citenamefont{Thakur, Chen, and
  Kapral}}]{thakur:11_1}
\bibinfo{author}{\bibfnamefont{S.}~\bibnamefont{Thakur}},
  \bibinfo{author}{\bibfnamefont{J.~X.} \bibnamefont{Chen}}, \bibnamefont{and}
  \bibinfo{author}{\bibfnamefont{R.}~\bibnamefont{Kapral}},
  \bibinfo{journal}{Angew. Chem. Int. Ed.} \textbf{\bibinfo{volume}{50}},
  \bibinfo{pages}{10165} (\bibinfo{year}{2011}).

\bibitem[{\citenamefont{Chen et~al.}(2018)\citenamefont{Chen, Chen, and
  Kapral}}]{CCK18}
\bibinfo{author}{\bibfnamefont{J.~X.} \bibnamefont{Chen}},
  \bibinfo{author}{\bibfnamefont{Y.~G.} \bibnamefont{Chen}}, \bibnamefont{and}
  \bibinfo{author}{\bibfnamefont{R.}~\bibnamefont{Kapral}},
  \bibinfo{journal}{Adv. Sci.} \textbf{\bibinfo{volume}{5}},
  \bibinfo{pages}{1800028} (\bibinfo{year}{2018}).

\bibitem[{\citenamefont{Desai and Kapral}(2009)}]{Desai_Kapral@Book:2009}
\bibinfo{author}{\bibfnamefont{R.~C.} \bibnamefont{Desai}} \bibnamefont{and}
  \bibinfo{author}{\bibfnamefont{R.}~\bibnamefont{Kapral}},
  \emph{\bibinfo{title}{Dynamics of self-organized and self-assembled
  structures}} (\bibinfo{publisher}{Cambridge University Press, New York},
  \bibinfo{year}{2009}).

\bibitem[{\citenamefont{Kapral and
  Showalter}(1995)}]{Kapral_Showalter@Book:1995}
\bibinfo{author}{\bibfnamefont{R.}~\bibnamefont{Kapral}} \bibnamefont{and}
  \bibinfo{author}{\bibfnamefont{K.}~\bibnamefont{Showalter}},
  \emph{\bibinfo{title}{Chemical waves and patterns}}
  (\bibinfo{publisher}{Kluwer, Dordrecht, The Netherlands},
  \bibinfo{year}{1995}).

\bibitem[{\citenamefont{Epstein and Pojman}(1998)}]{Epstein1998}
\bibinfo{author}{\bibfnamefont{I.~R.} \bibnamefont{Epstein}} \bibnamefont{and}
  \bibinfo{author}{\bibfnamefont{J.~A.} \bibnamefont{Pojman}},
  \emph{\bibinfo{title}{An Introduction to Nonlinear Chemical Dynamics}}
  (\bibinfo{publisher}{Oxford University Press, Oxford, UK},
  \bibinfo{year}{1998}).

\bibitem[{\citenamefont{Gaspard and
  Kapral}(2020)}]{Gaspard_Kapral@Research:2020}
\bibinfo{author}{\bibfnamefont{P.}~\bibnamefont{Gaspard}} \bibnamefont{and}
  \bibinfo{author}{\bibfnamefont{R.}~\bibnamefont{Kapral}},
  \bibinfo{journal}{Research} \textbf{\bibinfo{volume}{2020}},
  \bibinfo{pages}{9739231} (\bibinfo{year}{2020}).

\bibitem[{\citenamefont{Speck et~al.}(2015)\citenamefont{Speck, Menzel,
  Bialk{\'e}, and L{\"o}wen}}]{SMBL15}
\bibinfo{author}{\bibfnamefont{T.}~\bibnamefont{Speck}},
  \bibinfo{author}{\bibfnamefont{A.~M.} \bibnamefont{Menzel}},
  \bibinfo{author}{\bibfnamefont{J.}~\bibnamefont{Bialk{\'e}}},
  \bibnamefont{and}
  \bibinfo{author}{\bibfnamefont{H.}~\bibnamefont{L{\"o}wen}},
  \bibinfo{journal}{J. Chem. Phys.} \textbf{\bibinfo{volume}{142}},
  \bibinfo{pages}{224109} (\bibinfo{year}{2015}).

\bibitem[{\citenamefont{Selkov}(1968)}]{Selkov@EJB:1968}
\bibinfo{author}{\bibfnamefont{E.~E.} \bibnamefont{Selkov}},
  \bibinfo{journal}{Eur. J. Biochem.} \textbf{\bibinfo{volume}{4}},
  \bibinfo{pages}{79} (\bibinfo{year}{1968}).

\bibitem[{\citenamefont{Richter et~al.}(1981)\citenamefont{Richter, Rehmus, and
  Ross}}]{Richter_et_al@PTP:1981}
\bibinfo{author}{\bibfnamefont{P.~H.} \bibnamefont{Richter}},
  \bibinfo{author}{\bibfnamefont{P.}~\bibnamefont{Rehmus}}, \bibnamefont{and}
  \bibinfo{author}{\bibfnamefont{J.}~\bibnamefont{Ross}},
  \bibinfo{journal}{Prog. Theor. Phys.} \textbf{\bibinfo{volume}{66}},
  \bibinfo{pages}{385} (\bibinfo{year}{1981}).

\bibitem[{\citenamefont{Turing}(1952)}]{Turing:1952}
\bibinfo{author}{\bibfnamefont{A.~M.} \bibnamefont{Turing}},
  \bibinfo{journal}{Philosophical Transactions of the Royal Society B}
  \textbf{\bibinfo{volume}{237}}, \bibinfo{pages}{37} (\bibinfo{year}{1952}).

\bibitem[{\citenamefont{Castets et~al.}(1990)\citenamefont{Castets, Dulos,
  Boissonade, and Kepper}}]{Castets_et_al@PRL:1990}
\bibinfo{author}{\bibfnamefont{V.}~\bibnamefont{Castets}},
  \bibinfo{author}{\bibfnamefont{E.}~\bibnamefont{Dulos}},
  \bibinfo{author}{\bibfnamefont{J.}~\bibnamefont{Boissonade}},
  \bibnamefont{and} \bibinfo{author}{\bibfnamefont{P.~D.}
  \bibnamefont{Kepper}}, \bibinfo{journal}{Phys. Rev. Lett.}
  \textbf{\bibinfo{volume}{64}}, \bibinfo{pages}{2953} (\bibinfo{year}{1990}).

\bibitem[{\citenamefont{Ouyang and Swinney}(1991)}]{ouyang1991}
\bibinfo{author}{\bibfnamefont{Q.}~\bibnamefont{Ouyang}} \bibnamefont{and}
  \bibinfo{author}{\bibfnamefont{H.~L.} \bibnamefont{Swinney}},
  \bibinfo{journal}{Nature} \textbf{\bibinfo{volume}{352}},
  \bibinfo{pages}{610} (\bibinfo{year}{1991}).

\bibitem[{Not()}]{Note:Hopf}
\bibinfo{note}{The steady state solutions become unstable through a
  supercritical Hopf bifurcation for $\widetilde{k}_{+} \gtrapprox 0.0001$.}

\bibitem[{\citenamefont{Hull and White}(1990)}]{Hull_White@JFQA:1990}
\bibinfo{author}{\bibfnamefont{J.}~\bibnamefont{Hull}} \bibnamefont{and}
  \bibinfo{author}{\bibfnamefont{A.}~\bibnamefont{White}},
  \bibinfo{journal}{Journal of Financial and Quantitative Analysis}
  \textbf{\bibinfo{volume}{25}}, \bibinfo{pages}{87} (\bibinfo{year}{1990}).

\bibitem[{\citenamefont{Scheel}(1998)}]{Scheel@SJMA:1998}
\bibinfo{author}{\bibfnamefont{A.}~\bibnamefont{Scheel}},
  \bibinfo{journal}{SIAM Journal on Mathematical Analysis}
  \textbf{\bibinfo{volume}{29}}, \bibinfo{pages}{1399} (\bibinfo{year}{1998}).

\bibitem[{\citenamefont{Coullet et~al.}(1989)\citenamefont{Coullet, Gil, and
  Lega}}]{coullet1989}
\bibinfo{author}{\bibfnamefont{P.}~\bibnamefont{Coullet}},
  \bibinfo{author}{\bibfnamefont{L.}~\bibnamefont{Gil}}, \bibnamefont{and}
  \bibinfo{author}{\bibfnamefont{J.}~\bibnamefont{Lega}},
  \bibinfo{journal}{Phys. Rev. Lett.} \textbf{\bibinfo{volume}{62}},
  \bibinfo{pages}{1619} (\bibinfo{year}{1989}).

\bibitem[{\citenamefont{Straube et~al.}(2010)\citenamefont{Straube, Vermeer,
  Nicola, , and Mair}}]{gly-spiral}
\bibinfo{author}{\bibfnamefont{R.}~\bibnamefont{Straube}},
  \bibinfo{author}{\bibfnamefont{S.}~\bibnamefont{Vermeer}},
  \bibinfo{author}{\bibfnamefont{E.~M.} \bibnamefont{Nicola}}, ,
  \bibnamefont{and} \bibinfo{author}{\bibfnamefont{T.}~\bibnamefont{Mair}},
  \bibinfo{journal}{Biophys. J.} \textbf{\bibinfo{volume}{99}},
  \bibinfo{pages}{L04–L06} (\bibinfo{year}{2010}).

\end{thebibliography}



\end{document}